\documentclass[12pt,preprint]{aastex}
\newcommand{\be}{\begin{equation}}
\newcommand{\ba}{\begin{eqnarray}}
\newcommand{\ee}{\end{equation}}
\newcommand{\ea}{\end{eqnarray}}

\begin{document}
\title{Early Formation and Late Merging of the Giant Galaxies}

\author{Liang Gao\altaffilmark{1} Abraham Loeb\altaffilmark{2}
P. J. E. Peebles\altaffilmark{3} Simon D. M. White\altaffilmark{1} and
Adrian Jenkins\altaffilmark{4}} 
\altaffiltext{1}{Max-Planck-Institut f{\"u}r Astrophysik, D-85748
Garching bei M\"unchen, Germany}
\altaffiltext{2}{Astronomy Department, Harvard University, Cambridge
MA 02138, USA} 
\altaffiltext{3}{Joseph Henry Laboratories, Princeton University, 
Princeton NJ 08544, USA} 
\altaffiltext{4}{Institute for Computational Cosmology, Physics
Department, University of Durham, Durham DH1 3LE, UK}
\begin{abstract}
The most luminous galaxies in the present Universe are found at the
centers of the most massive dark matter haloes, rich galaxy clusters.
In the $\Lambda$CDM cosmology, such massive halo cores are present at
redshift $z=6$ with a comoving number density (as a function of mass
interior to $\sim 10$~kpc) that is comparable to today's value.  The
identity of the matter in these central regions is, however, predicted
to change as major mergers bring together stars and dark matter from
initially well separated sub-units. We use N-body simulations to investigate
how these mergers push pre-existing matter outwards in the dominant
galaxy while preserving the inner density profile of collisionless
matter. It appears that the central regions of large galaxies end up
dominated by stars formed in a number of dense cores, well before the
last major mergers. The density profile of collisionless matter (stars
and dark matter combined) in these central regions appears to be
stable and to have attractor-like behavior under merging. This
suggests that the baryon loading associated with dissipative
contraction and star formation may be erased as subsequent mergers
drive the mass distribution back to a universal profile. Such
suppression of the effects of baryon loading, along with the early
assembly of mass concentrations, may help resolve some apparent
challenges to the CDM model for structure formation.
\end{abstract}


\section{Introduction}

Recent merger-driven evolution of the most massive galaxies was under
discussion well before the introduction of the Cold Dark Matter (CDM)
model for structure formation (see for example Toomre \&\ Toomre 1972,
\S  VII.b; Ostriker \&\ Tremaine 1975), and has long been recognized
as an important process within the CDM model (Frenk et al. 1985).  A
less widely discussed aspect of this model is that dark matter halos
with characteristic velocities and comoving number densities
characteristic of the luminous parts of large galaxies form at
redshifts well above unity (Loeb \&\ Peebles 2003). The dichotomy --
very significant events in the history of the massive galaxies at low
and high redshift -- may be mirrored in the observational data: there
is clear evidence for merging and evolution beyond aging of the star
populations at redshift $z<1$, and clear evidence also for the
presence of giant galaxies with old star populations at redshifts well
above unity (Conselice et al., 2003).

We discuss the relation between these two aspects of galaxy formation
in the CDM model by combining arguments based on analytic fitting
functions and on direct numerical N-body simulations. The early
formation of mass concentrations similar to those in the luminous
parts of the most massive present-day galaxies is reviewed in \S 2. In
\S 3 we present $\Lambda$CDM simulations which have
sufficient resolution to follow the assembly of the regions which
house the central dominant galaxy in observed rich clusters. The
details of this assembly are analyzed in \S 4.  Mergers among massive
halos at redshifts between 0.5 and 4 bring more matter into the
innermost $10$~kpc than remains from the dominant progenitor at higher
redshift. Logical and observational consistency with the early
formation of massive systems leads to three conditions. First, most of
the matter present in the centers of the dominant halos at $z=6$ has
to be displaced outwards during mergers. We show this effect in the
simulations.  Second, the hierarchy of mergers has to preserve the
stellar concentration within radii $\sim 10$~kpc. This may reflect the
fact that in merger simulations the dense regions (where stars seem
most likely to form) tend to end up in the dense regions of the merger
remnant. We present in \S 4 a statistical measure that illustrates
this preservation effect. Third, the characteristic density profile of
a virialized halo of collisionless matter has to be stable under a
sequence of disturbances from major mergers.  As discussed in \S 4.3,
this attractor effect is supported by the simulations.  An important
observational consequence may be the suppression of the adiabatic
baryon loading associated with gas cooling and star formation. 

Our central conclusion is that in the $\Lambda$CDM
cosmology giant galaxies exist at redshift $z=3$ with close to the
present comoving number density, in terms of the total mass
measured within physical radius $r\sim 10$~kpc. At this time they
may have up to half the present star mass in this region. This
would be quite different from the indications from at least some
semi--analytic models for galaxy formation (e.g. Baugh et al. 1998,
figure 13), but in line with a considerable variety of  -- though not
all -- observational indications (as reviewed in Peebles
2002). Further considerations on whether our interpretation of the
$\Lambda$CDM model agrees with the observations are presented in \S 5.

Throughout this paper, we assume the standard $\Lambda$CDM cosmological
parameters $\Omega_m=0.3$, $\Omega_\Lambda=0.7$, $\Omega_b=0.04$,
$\sigma_8=0.9$, $n=1$, and Hubble constant $H_0=100 h~{\rm
km~s^{-1}~Mpc^{-1}}$ with $h=0.7$.

\begin{figure}[htbp]
\plotone{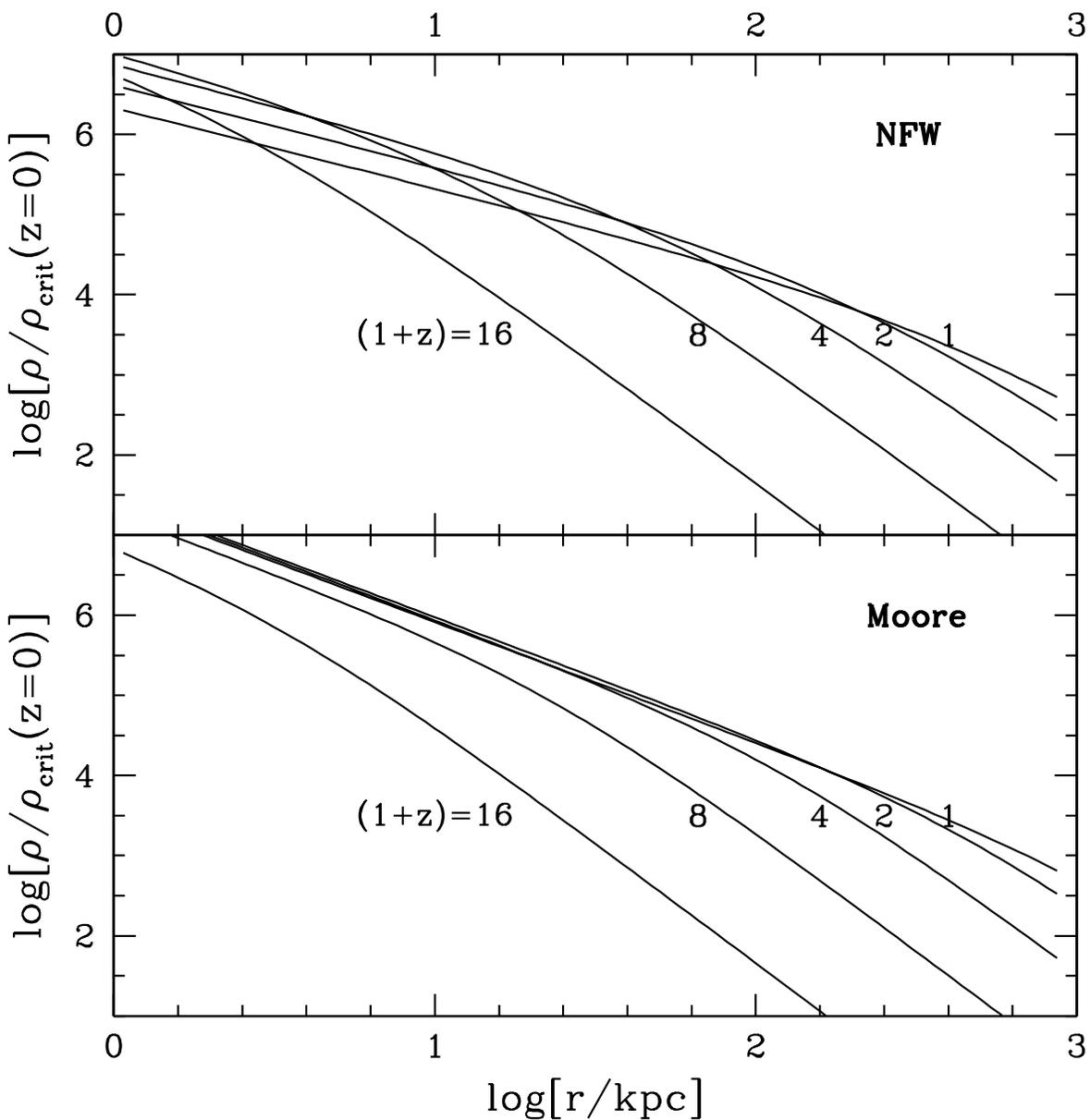}
\caption{Density runs at various redshifts for halos with comoving density
$n(>M)=10^{-7}~{\rm Mpc}^{-3}$. Physical rather than comoving units
are used both for the radius and for the density.}
\label{fig-1}
\end{figure}

\section{Formation of Mass Concentrations Characteristic of the
Most Massive Galaxies}

Analytic fitting functions can be combined with analytic formulae for
halo abundance to make $\Lambda$CDM predictions for the characteristic
mass density run in rare, very massive halos.  In Figure 1 we show
results for halos assumed to have a fixed comoving number density,
corresponding to physical density $n=10^{-7}a(t)^{-3}~{\rm Mpc^{-3}}$;
the profiles are plotted at redshifts corresponding to factor of two
steps in the cosmological expansion factor $a=(1+z)^{-1}$.  The most
striking impression from this plot is how little the mass distribution
changes in the inner regions after $1+z=8$.  Fukushige \& Makino
(2001) were led by their numerical simulations to 
propose that that the run of density in physical units in the inner
power law part is approximately independent of time, a behavior
suggested previously by the simple theoretical model of Syer \& White
(1998) for the assembly of halos through merging.  Loeb \& Peebles
(2003) were independently led to the same proposal from the fitting
function analysis in Figure 1.

This figure, computed as described in Loeb \& Peebles (2003), is based on
the Press-Schechter mass function (including the modification by Sheth
\& Tormen 1999; see also Sheth, Mo \&\ Tormen 2001) and analytic density
profiles.  The Navarro, Frenk, \& White (1997, hereafter NFW) shape
with concentration parameter $c=4$ is used in the top panel, while the
Moore (Moore et al. 1999; Ghigna et al. 2000) profile with
concentration $c=4/1.72$ is used in the bottom panel. (See Klypin et
al. 2001 for the conversion factor, 1.72, between the two profiles.)
Clearly, the mass in the inner $10 h^{-1}$~kpc of these rare halos is
predicted to evolve very little for $z\la 6$. That is, according to the
$\Lambda$CDM model, massive cores similar to those which house the
largest present-day galaxies already existed just one billion years
after the Big Bang. We will see in the next section that our
simulations of $\Lambda$CDM bear out this result from the fitting functions.

\section{High Resolution Simulations of Massive Halo Assembly}

The numerical results in this paper are based on a set of 8
simulations of the formation of a massive galaxy cluster halo in our
standard $\Lambda$CDM model. These 8 halos, which are part of the suite
of simulations analysed in Navarro et al (2003), range in virial mass
between $4.5\times 10^{14}h^{-1}{\rm M_\odot}$ and $8.5 \times
10^{14}h^{-1}{\rm M_\odot}$.  They are chosen from a simulation of a
representative cubic region of side $479h^{-1}$Mpc (the VLS simulation
of the Virgo Consortium, see Jenkins et al. 2001 and Yoshida Sheth \&\
Diaferio 2001), which contains 41 halos with mass exceeding $4.5
\times 10^{14}h^{-1}{\rm M_\odot}$.  Our objects thus have an
effective abundance of $3.7\times 10^{-7}h^3{\rm Mpc^{-3}}$. This is
the observed present-day abundance of galaxies with luminosity greater
than $8L_*$. Almost all such systems are indeed the central dominant
galaxies within rich clusters.

We resimulated each of our 8 halos, as in Navarro, Frenk \& White
(1997), with greatly improved resolution in the cluster and its
immediate environment and with degraded resolution outside this
region. The mass of an individual dark matter particle in the high
resolution region is $5.12 \times 10^8h^{-1}{\rm M_\odot}$ and the
gravitational softening parameter is $5.0h^{-1}$~kpc in comoving
units. The simulations were carried out with the publicly available
parallel N-body code GADGET (Springel, Yoshida \& White 2001).

We show images of the evolution of the mass distribution in these 8
halos in Figure~2. The three sets of panels show the halo material at
three different redshifts, $z=0, 1$ and 3. Each panel is $5h^{-1}$Mpc
across in physical (not comoving) units. Each shows only the matter
which is within $r_{200}$ of the cluster center at $z=0$, so that the
same particles are used to make corresponding images in each of the
three sets. As usual, we define $r_{200}$ to be the radius within
which the mean enclosed density is 200 times the critical value.  It
is striking that although all the halos are centrally concentrated and
relatively regular at $z=0$, the material which makes them up was in
all cases in several disjoint and well separated pieces at $z=1$ and
was in many pieces at $z=3$.

In the images in Figure~2 the particles which lie within
$10~h^{-1}$~kpc of halo center at $z=0$ are shown in black in all three
sets. It is remarkable that in all cases these particles also come
from several different objects at $z=3$.  The same is true even at
$z=1$ in many cases. We analyse the details of core assembly in
more detail in the next section.

The stability of the central mass concentrations predicted in \S 2 can
be seen directly in these simulations. Figure~3 shows the mass
within physical radius $r=10h^{-1}$~kpc around the center of the most
massive progenitor of the final halo at discrete time steps and in
each of the 8 simulations. Notice that the vertical axis is linear in
these plots. The variations in mass are relatively small and show no 
consistent trend for $a > 0.15$, corresponding to $z<6$. This is in good
agreement with  Figure~1. That is, the CDM model predicts that at
$z<6$ there is little evolution of the mass within a radius
characteristic of the luminous parts of the largest galaxies.
Note, however, that the object plotted in each panel is {\it not} the
same at each time: the most massive progenitor of a cluster at
$z=6$ does not necessarily evolve into its most massive progenitor at
$z=4$ which may not evolve into its most massive progenitor at
$z=2$. We indicate this effect in the plots; working back from $z=0$,
we toggle the plotting symbol between filled and open each time the
most massive progenitor changes identity.

\begin{figure}[htbp]
\plotone{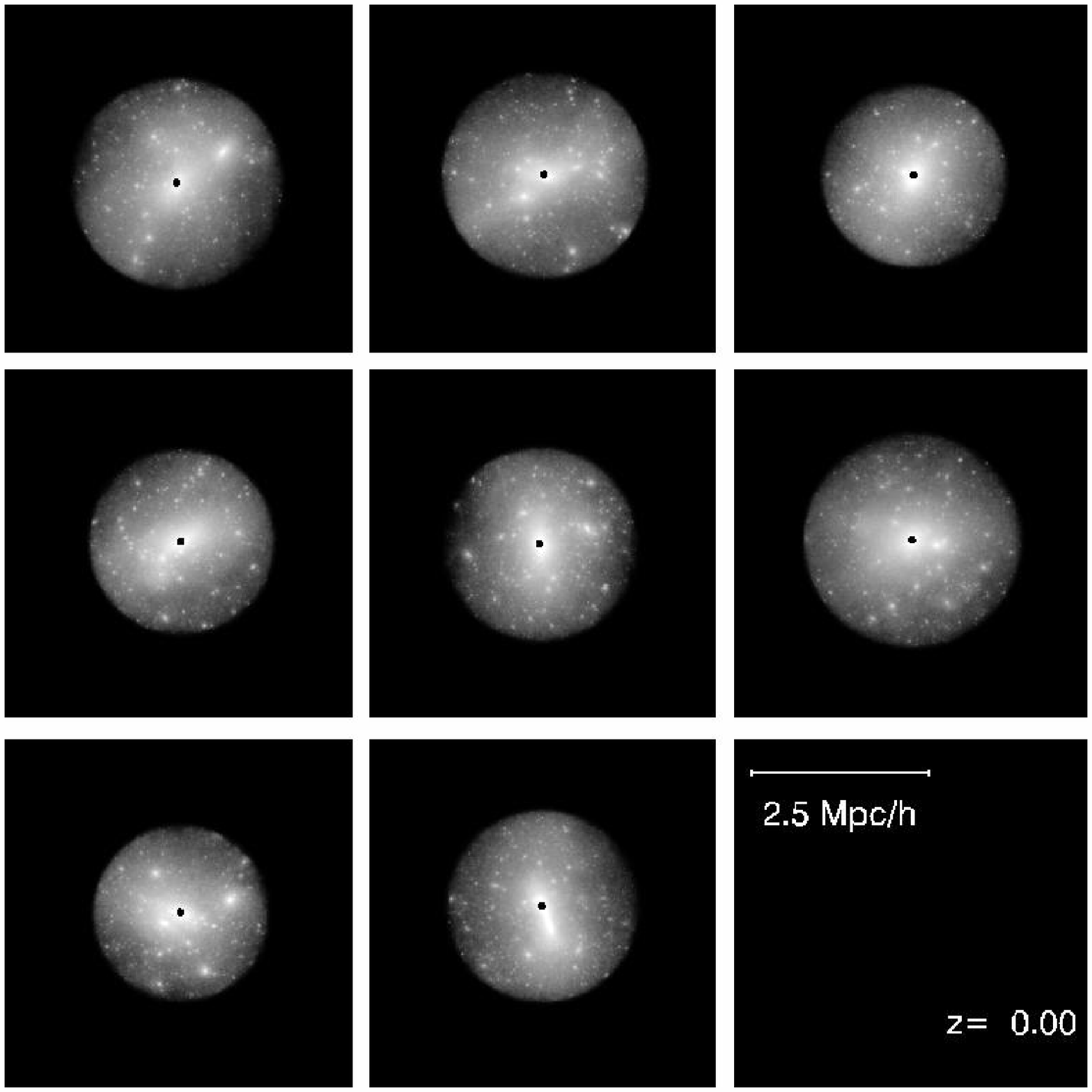}
\label{fig-2a}
\end{figure}

\begin{figure}[htbp]
\plotone{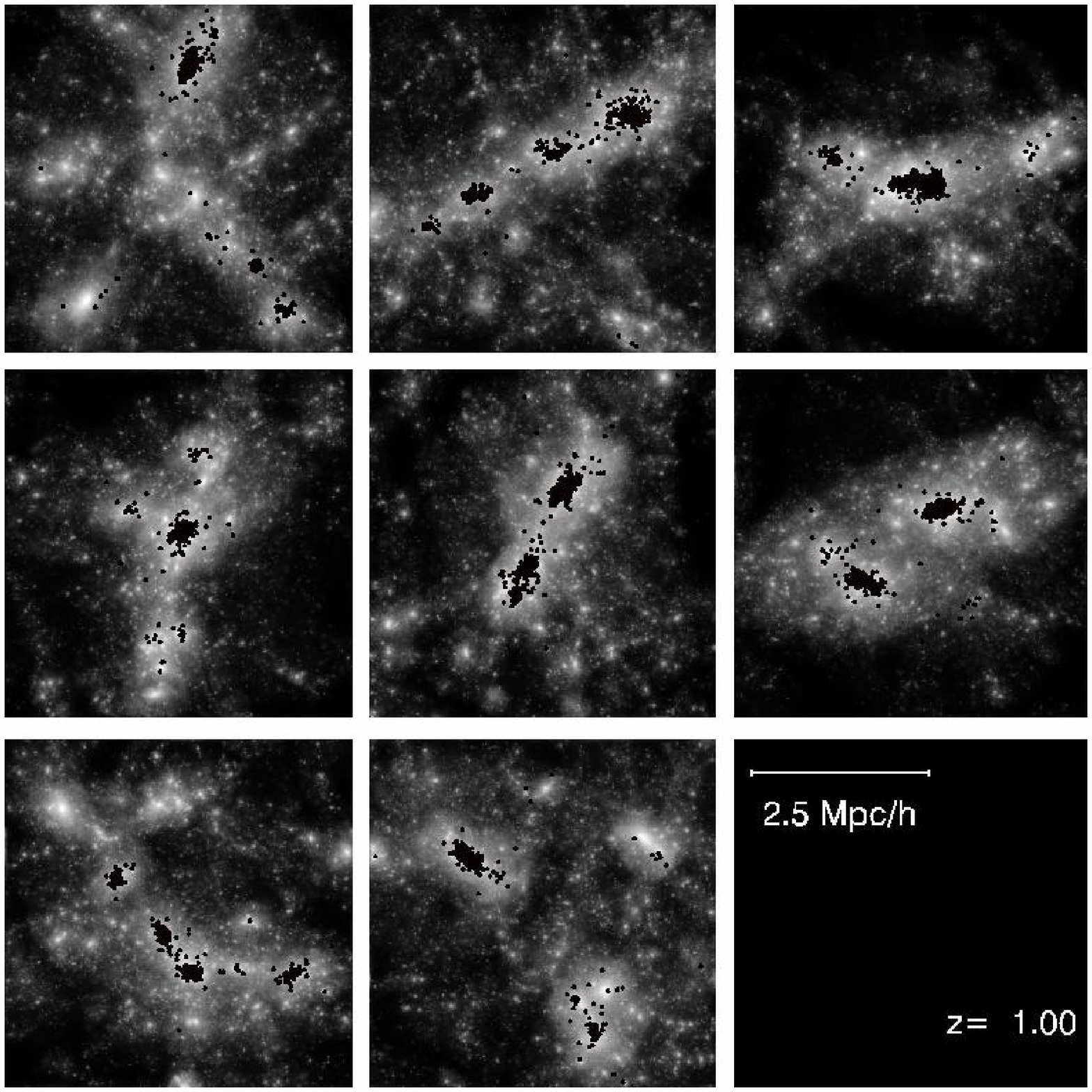}
\label{fig-2b}
\end{figure}

\begin{figure}[htbp]
\plotone{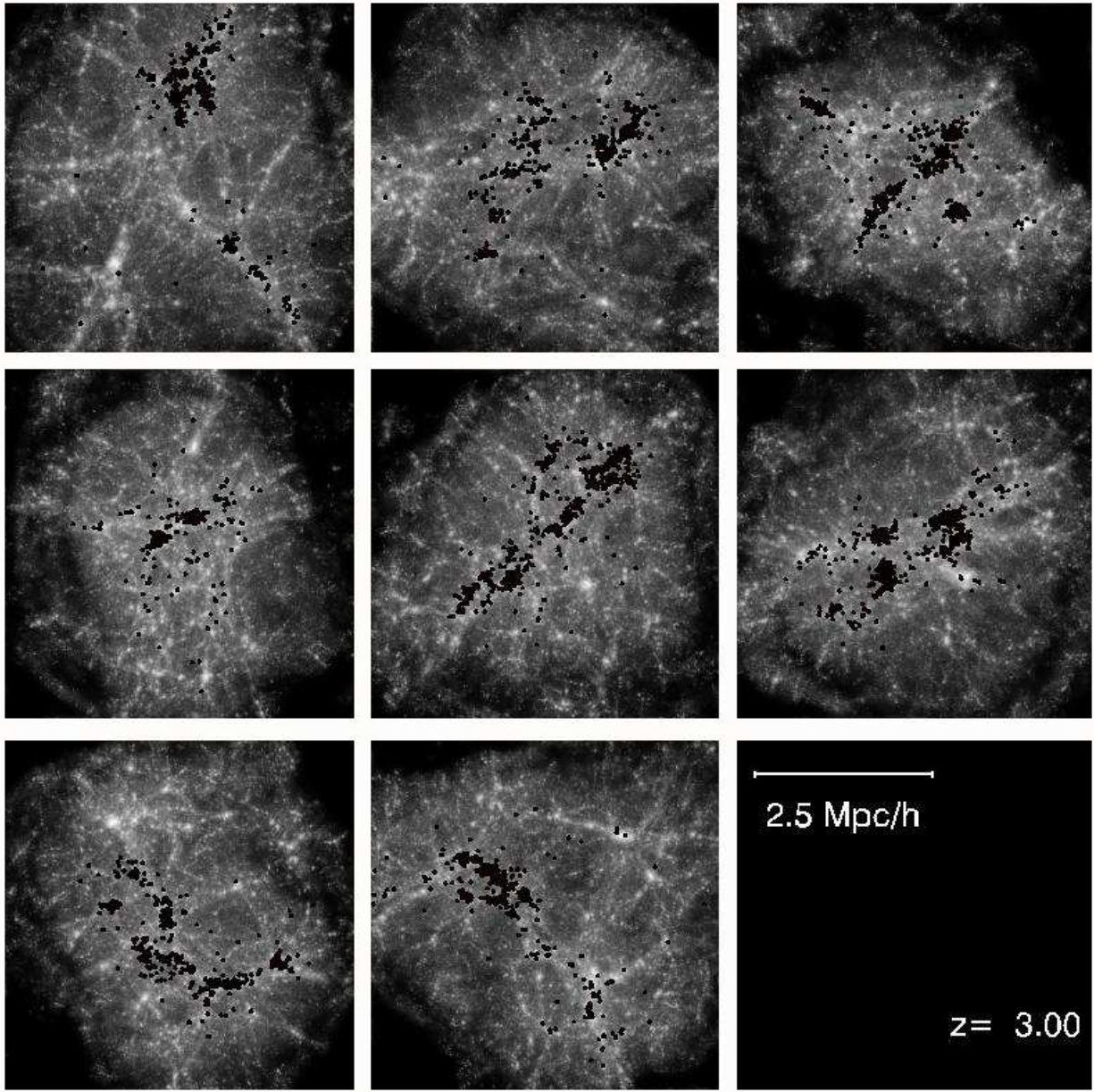}
\label{fig-3b}
\caption{Images of the mass distribution at $z=0, 1$ and $3$ in our 8
simulations of the assembly of cluster mass halos. Each plot shows
only those particles which lie within $r_{200}$ of halo center at
$z=0$.  Particles which lie within $10 h^{-1}$~kpc of halo center at this time
are shown in black. Each image is $5h^{-1}$Mpc on a side in physical
(not comoving) units.}
\label{fig-2}
\end{figure}

\begin{figure}[htbp]
\plotone{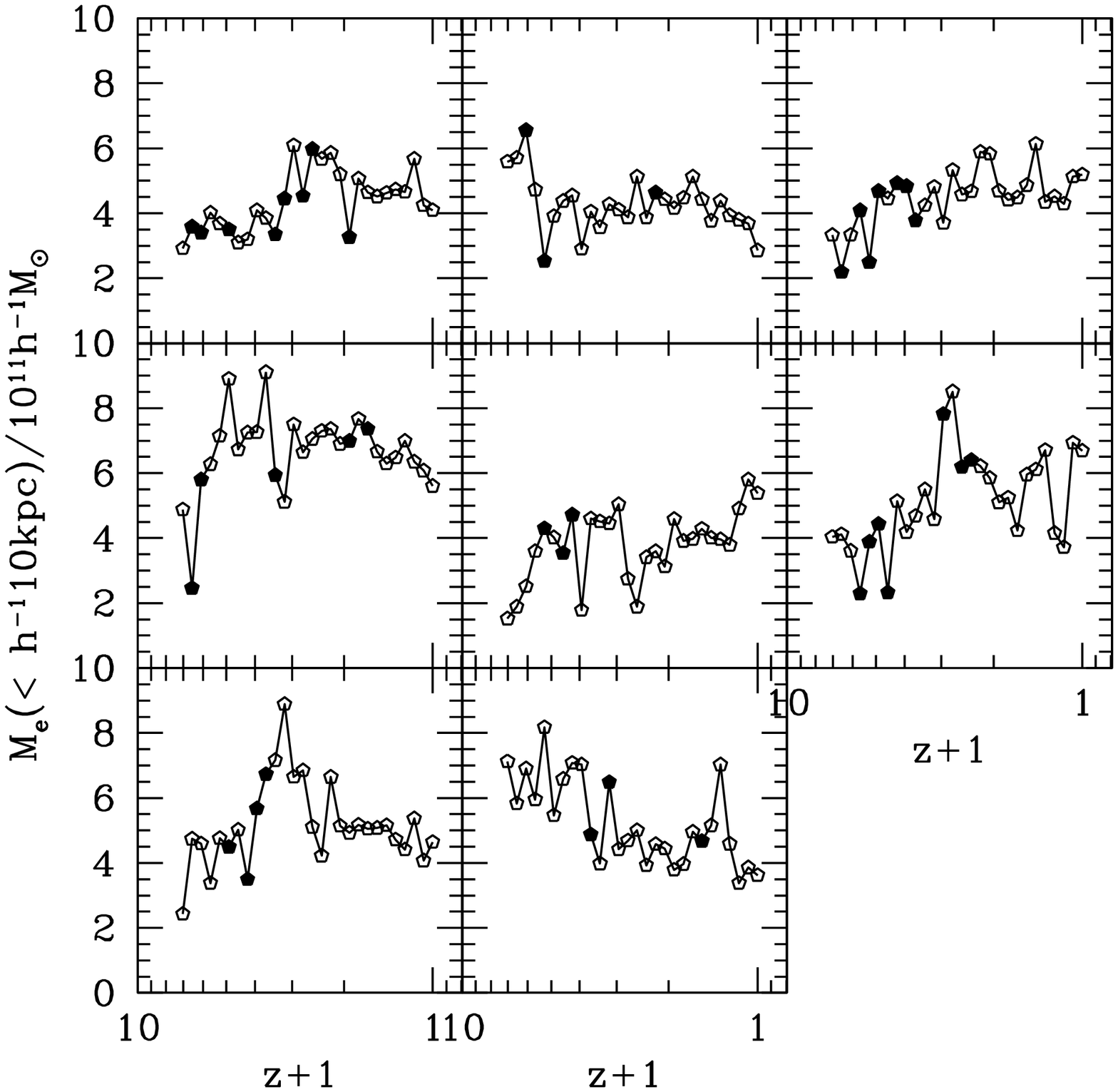}
\caption{The total mass within physical distance $10 h^{-1}$~kpc of
the center of the most massive progenitor of the final halo at each
time plotted and for each of our 8 simulations. Symbols switch between
filled and open each time the identity of the most massive progenitor
changes. }
\label{fig-3}
\end{figure}

\section{Mergers and Relaxation at Low Redshifts}

In this section, we consider the predicted rearrangement of matter in the
cores of pure CDM halos at low redshift, and then discuss why the stars in
giant galaxies might be expected to remain concentrated in the centers of
the halos as observations require. Finally, we consider the idea that
the net mass distribution in stars plus dark matter, both considered as
collisionless particles, tends to relax toward the NFW form.

\subsection{Rearrangements of the Dark Matter}
Loeb \& Peebles (2003) discuss the evolution of the halo structure shown in
Figure~1 in terms of an ``inside-out'' growth process, whereby mass is
added to galaxies in ``onion shells'' with declining density as a function
of cosmic time. This can indeed reproduce the behavior in Figure~1,
but cannot be the entire story because, as Figure~2 shows, late
mergers add material even to the very center of the main halo and so
must affect the distribution of the other matter there.

Figure~4 makes this point more quantitatively. We identify the
particles which are within 10$h^{-1}$kpc of the center of each cluster
halo at $z=0$, and we then follow them back in time.  The circles in
each panel show the fraction of these particles which are already
within $100 h^{-1}$~kpc (physical) of the center of their dominant
concentration at each earlier redshift. (We identify the center of
this dominant concentration by calculating the gravitional potential
of each particle in the set due to all the others, and then choosing
the most bound particle.) Note that the dominant concentrations used to
make this plot are often not the most massive progenitors which were
used to make Figure~3. Both figures illustrate the point that, in the
$\Lambda$CDM model, mergers at low redshifts have a substantial effect
on the innermost regions of large halos. Only $20$--$50\%$ of the mass
that now lies within $10h^{-1}$~kpc of the center of a massive halo
was closer than $100h^{-1}$~kpc to their dominant concentration at $z
= 6$, and typically no more than $50\%$ was closer than $100
h^{-1}$~kpc at $z= 2$. The rest of the mass was added to the cores by
late mergers. These major mergers are visible in Figure~4 as abrupt
changes in $F(z)$ which are often accompanied by large fluctuations in
the mass within $10h^{-1}$~kpc.

\begin{figure}[htbp]
\plotone{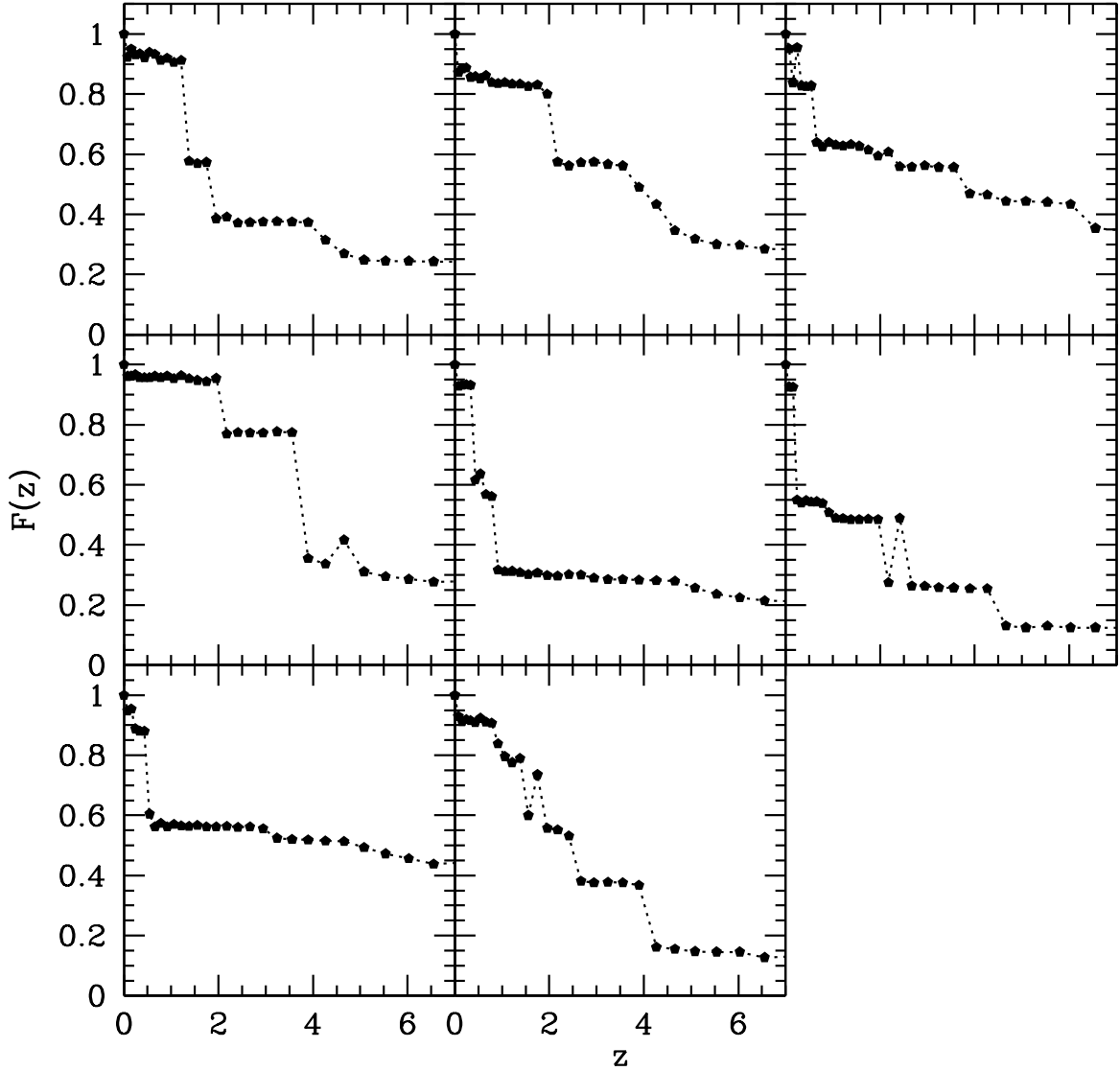}
\caption{History of addition of the matter now in the central parts of
massive halos. The black curves show the fraction of the particles
at $r < 10 h^{-1}$~kpc at $z=0$ which lie within $100 h^{-1}$~kpc
(physical) distance from the center of their main concentration at
each earlier redshift $z$.}\label{fig-4}
\end{figure}

The matter present in the central $10 h^{-1}$~kpc of each massive
concentration at high redshift must have been displaced to make room
for the matter subsequently added by mergers. We illustrate this
process in Figure~5.  We start by selecting all particles within $10
h^{-1}$~kpc (physical) of the center of the most massive progenitor of
each halo at $z=6$. Since many of these particles have apogalactica
well outside $10 h^{-1}$~kpc, we plot their cumulative radial
distributions at $z=5.5$ after they have had time to phase-mix around
their orbits. We then identify this same set of particles at a series
of later times and plot the cumulative radial distribution about the
center of their dominant concentration.  (This center is defined as
the most bound particle of the set, as above.) One sees a systematic
trend for these distributions to broaden with time, the median
distance typically increasing by a factor of about two from $z = 4$ to
the present. Notice, however, that in two of the eight cases the
dominant concentration of these particles at $z=0$ is {\it not} at the
center of the main halo, but at the center of one of its more massive
subhalos.

The late assembly of the matter which does finally end up at halo
center is illustrated by the complementary plot in Figure~6. Here we
again select all particles which are within $10 h^{-1}$~kpc of halo
center at $z=0$ and then plot cumulative radial distributions about
the center of their dominant concentration at a series of earlier
times.  (These are the same particle sets and center definitions used
to make Fig.~4.)  We plot the lowest redshift curves for $z=0.07$
rather than for $z=0$ in order to show a properly phase-mixed,
quasi-equilibrium distribution.  There is little evolution subsequent
to $z=0.33$ in 7 cases, subsequent to $z=1$ in 3 cases, and subsequent
to $z=2$ in one case. At higher redshifts, however, substantial
fractions of the particles are further than $100 h^{-1}$~kpc from
center of the dominant concentration in all cases.  This behavior
reflects the late addition of matter to the cores of the galaxies, as
already illustrated in Figure~4.

\begin{figure}[htbp]
\plotone{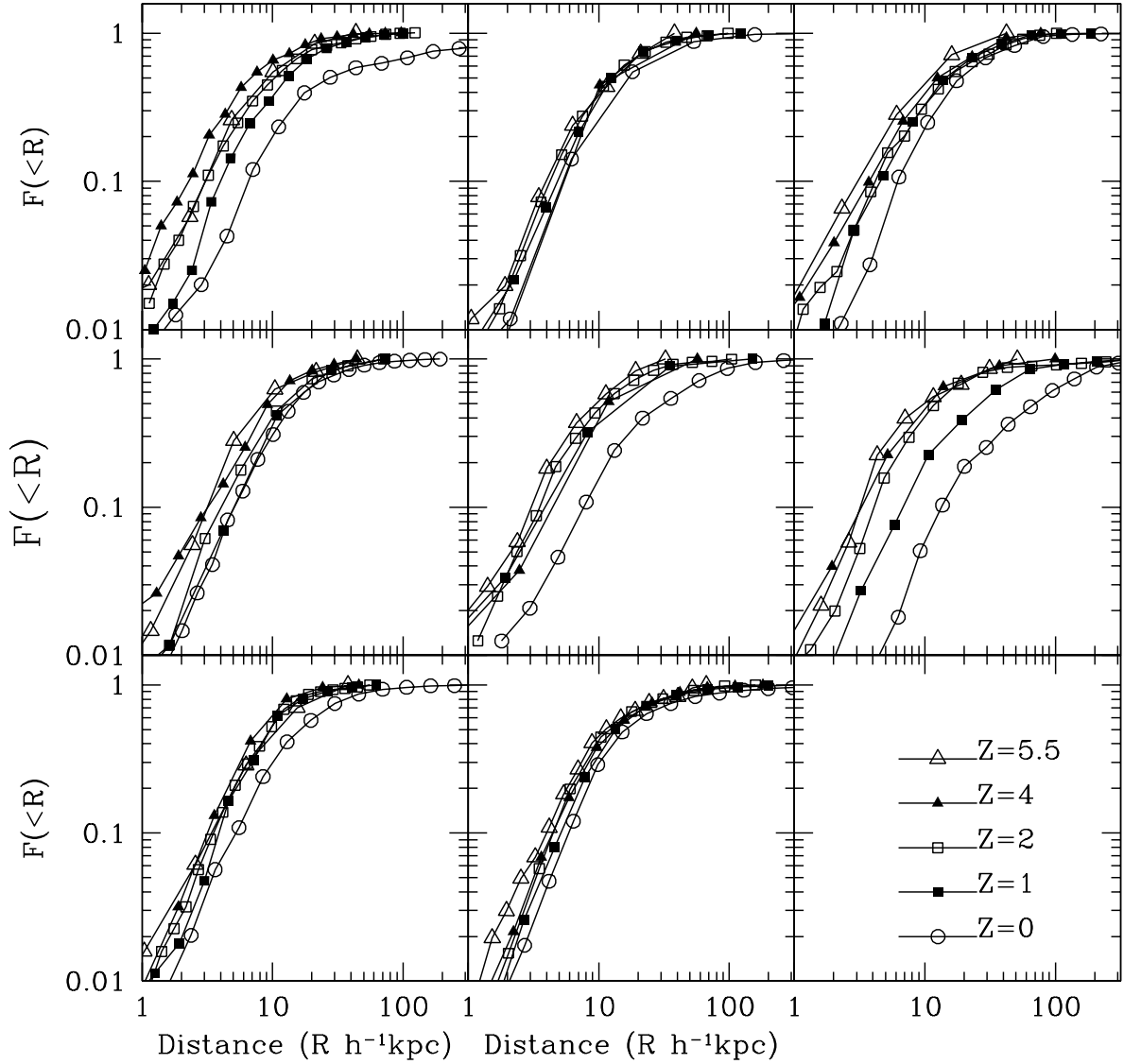} \caption{Cumulative radial distributions at a series
of later redshifts for the particles that were within $10 h^{-1}$~kpc
of the center of the most massive $z=6$ progenitor of each cluster
halo.  Distances are all in physical units and are measured from the
center of the dominant concentration of each particle set at each
redshift. Note that for the middle clusters in the top and bottom rows
(numbers C2 and C8) this dominant concentration does not coincide with
the cluster center at $z=0$ but with one of the more massive
substructures.} \label{fig-5}
\end{figure}

\begin{figure}[htbp]
\plotone{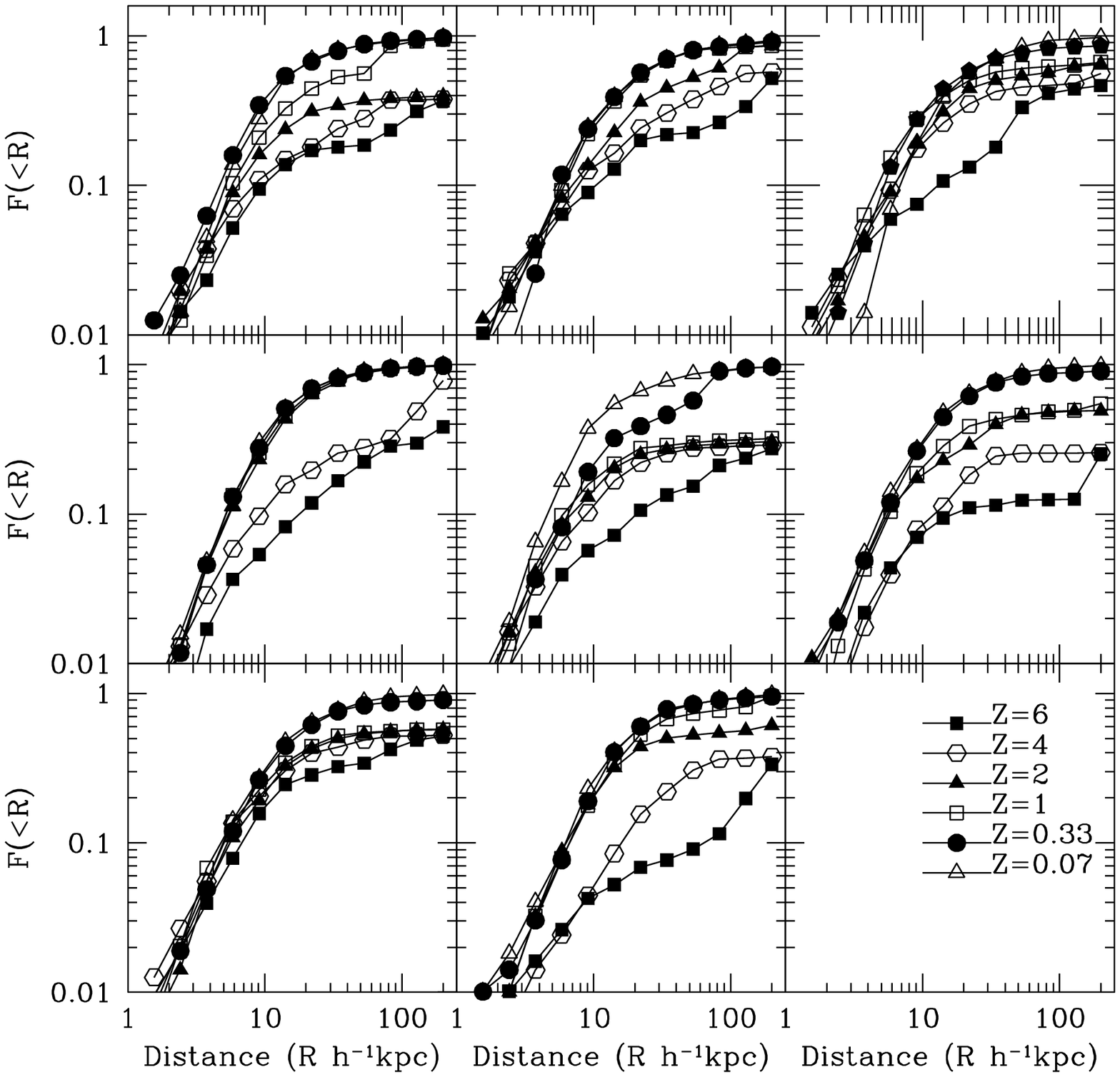}
\caption{Evolution of cumulative radial distributions, as in Figure~5,
but now for particles which are within $10 h^{-1}$~kpc of halo center
at the final time, $z=0$. These are same particle sets (with the same
definition of density center) already used to make Fig.~4. }
\label{fig-6}
\end{figure}

\subsection{The Distributions of Stars}

The star populations in giant ellipticals are typically old (a
familiar and well established observation, as evidenced by the
discussions by Oke 1971, 1984 and Hamilton 1985; for recent data see
Bernardi et al. 2003). Once formed, stars behave dynamically as 
collisionless matter. Since stars make substantial contributions to
the mass within the half-light radii $r_e\la 10 h^{-1}$~kpc of massive 
galaxies (Romanowsky et al. 2003, and references therein)
merger-driven rearrangements of matter must not have substantially
diluted the central concentrations of stars by the addition of
nonbaryonic dark matter. A full analysis of the predicted effect of
dilution is beyond the scope of this paper -- and perhaps
beyond what is now computationally feasible -- but we can offer two
simple relevant considerations. First, the condition that dilution is
modest is in line with the familiar tendency in numerical simulations
for the dense parts of merging halos to end up in the dense parts of
the merger remnant (White 1980; Barnes 1992; Dubinski 1998).

The second consideration is based on the same sets of particles already
used in Figures~4 and 6, namely those particles that
are within $r<10 h^{-1}$~kpc of the center of each dominant halo at
$z=0$. Figure~7 shows the evolution with redshift of the cumulative
distribution of ambient physical density around each of these
particles, estimated by means of a standard SPH spline kernel which
averages over the positions of the 25 nearest neighbors. Note that the
final time shown is $z=0.07$ rather than $z=0$ so that the particle
distribution is properly phase-mixed.
The median ambient density around this particle set increases by a
factor of about six from $z=6$. This is a result of our selection
procedure, which preferentially picks out those particles which have
been scattered into the most strongly bound orbits by 2--body effects
and by the violent relaxation which accompanies merging. The median
ambient density for these particles at $z=6$ is typically about 
$5\times 10^{6}M_\odot {\rm kpc}^{-3}$, which is 500 times the mean 
density at that epoch. The matter now in the central regions of a
giant galaxy was thus already in the inner regions of virialised
objects at $z=6$, and hence could have experienced substantial 
star formation at that time.

This mixing process is explored in a different way in Figure~8.  Among
all the particles that lie within $r_{200}$ in each final cluster we
identify the 1000 which have the largest ambient density at $z=6$. We
then plot cumulative ambient density distributions for these particle
sets at lower redshifts beginning with $z=5.53$. These distributions
broaden with time as relaxation scatters particles into lower density
regions. At $z=0$ their median ambient density is typically $3\times
10^{6}{\rm M_\odot} {\rm kpc}^{-3}$, which is a factor 9 smaller than
at $z=5.5$ but still  $10^5$ times the present cosmic
mean density. Note that much of this broadening occurs between $z=1$
and $z=0$, and is actually a consequence of 2--body scattering.  In
simulations of even better mass resolution, we would expect the
reduction in density at late times to be significantly lower. It is
important to realise, however, that not all these ``dense'' particles
from high redshift end up in the central ``galaxy''. Typically about
40\% of them lie within $100 h^{-1}$~kpc of the center of the final
halo; most of the others lie near the center of one of its
substructures. If we consider these particles to represent the matter
which was already illuminated by star formation at $z=6$, then the
corresponding light is today associated both with the dominant central
galaxy in each halo and with other cluster galaxies. 

A comparison of the distribution of matter that is illuminated now in
the giant galaxy with that which was plausibly already illuminated at
$z = 6$ is presented in Figure~9. This shows, for the particles used
in Figure~8, the present cumulative radial distribution about the
center of the final halo. In 6 of our 8 halos the largest single
concentration of these ``early dense'' particles is in the central
object, with 20\% to 50\% within $100h^{-1}$~kpc of halo center. In
the remaining two objects, however, the dominant concentrations are in
subhalos offset by 150 to 500$h^{-1}$~kpc from the center of the main
halo, so that the bulk of the earliest stars are predicted to be
in non-central galaxies.

In these rare massive halos at $z=6$ the virial radius (at density
contrast $\sim 200$) is comparable to the half-light radius $r_e \sim
10h^{-1}$~kpc of the bright galaxy at the center of the present day
descendant. If most of the baryons then within this region had
promptly collapsed to stars, the stellar mass fraction within
$r=10h^{-1}$~kpc at $z\sim 6$ would have been about equal to the
primeval mass fraction, that is, about one fifth of the total mass
within the present half-light radius. The remaining $\sim 80$ percent
of the stars would have been added later, by merging with other
concentrations of generally old stars.  Roughly in line with this, the
indication from Figure~3 is that $\sim 20$--$50\%$ of the mass now
interior to $r=10h^{-1}$~kpc was added at $3\la z\la 6$, and about
half of the mass was added at $z<3$. We must assume that most of the
added mass was stellar, so that the core can be star-dominated
today. If the mass displaced out of this radius were primarily CDM,
the stellar mass interior to $r_e\sim 10h^{-1}$~kpc would have roughly
doubled since $z\sim 3$. Since $M(<r)\propto r^{\beta}$ with $\beta
\sim 1$ in the core, the effective radius $r_e$ of starlight would
have about doubled since $z\sim 3$.

\begin{figure}[htbp]
\plotone{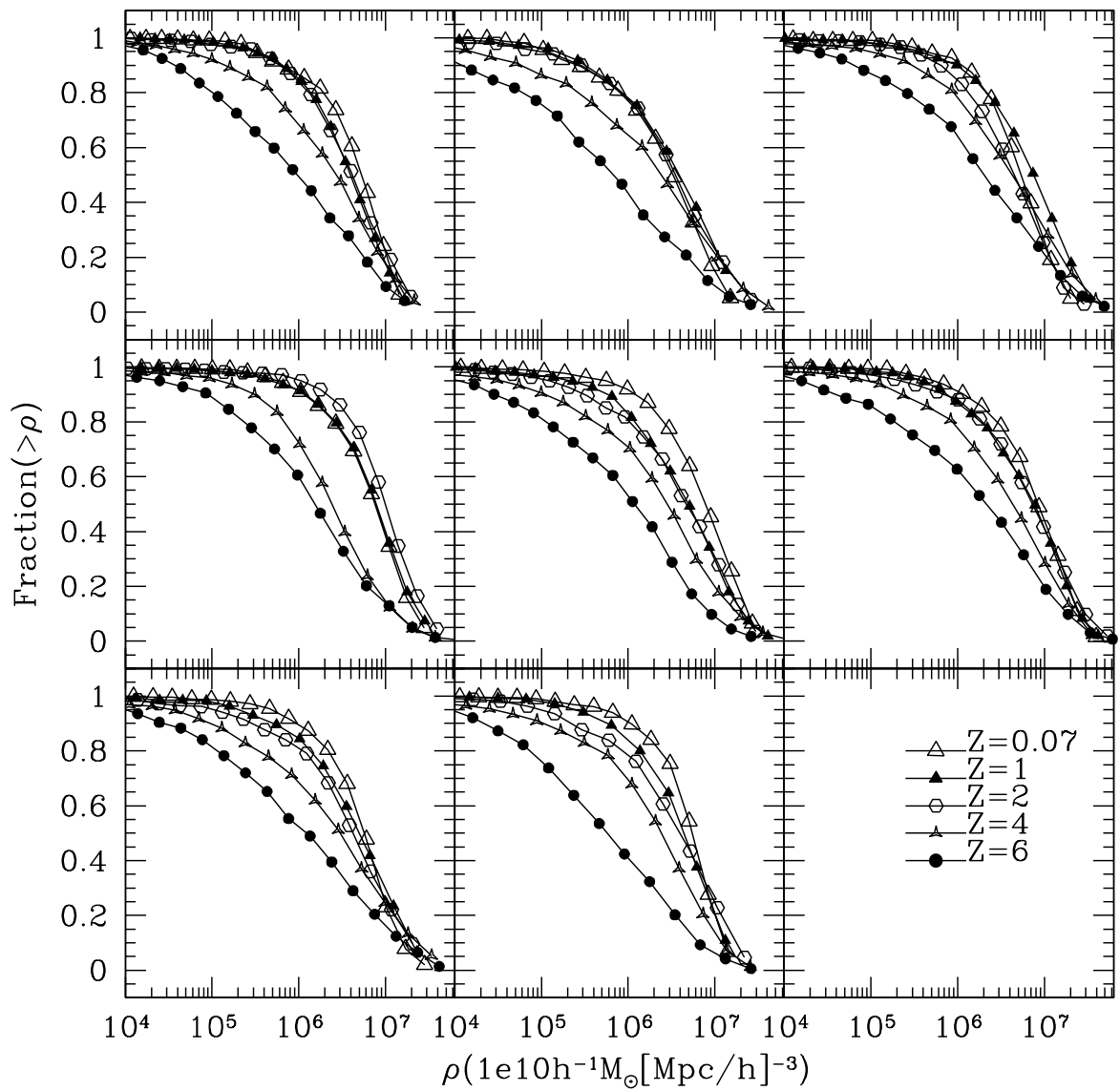} \caption{Evolution of the cumulative distribution
of ambient physical density for particles which lie within $10
h^{-1}$~kpc of halo center at $z=0$.  These are the same particle
sets used to make Figures 4 and 6.} \label{fig-7}
\end{figure}

\begin{figure}[htbp]
\plotone{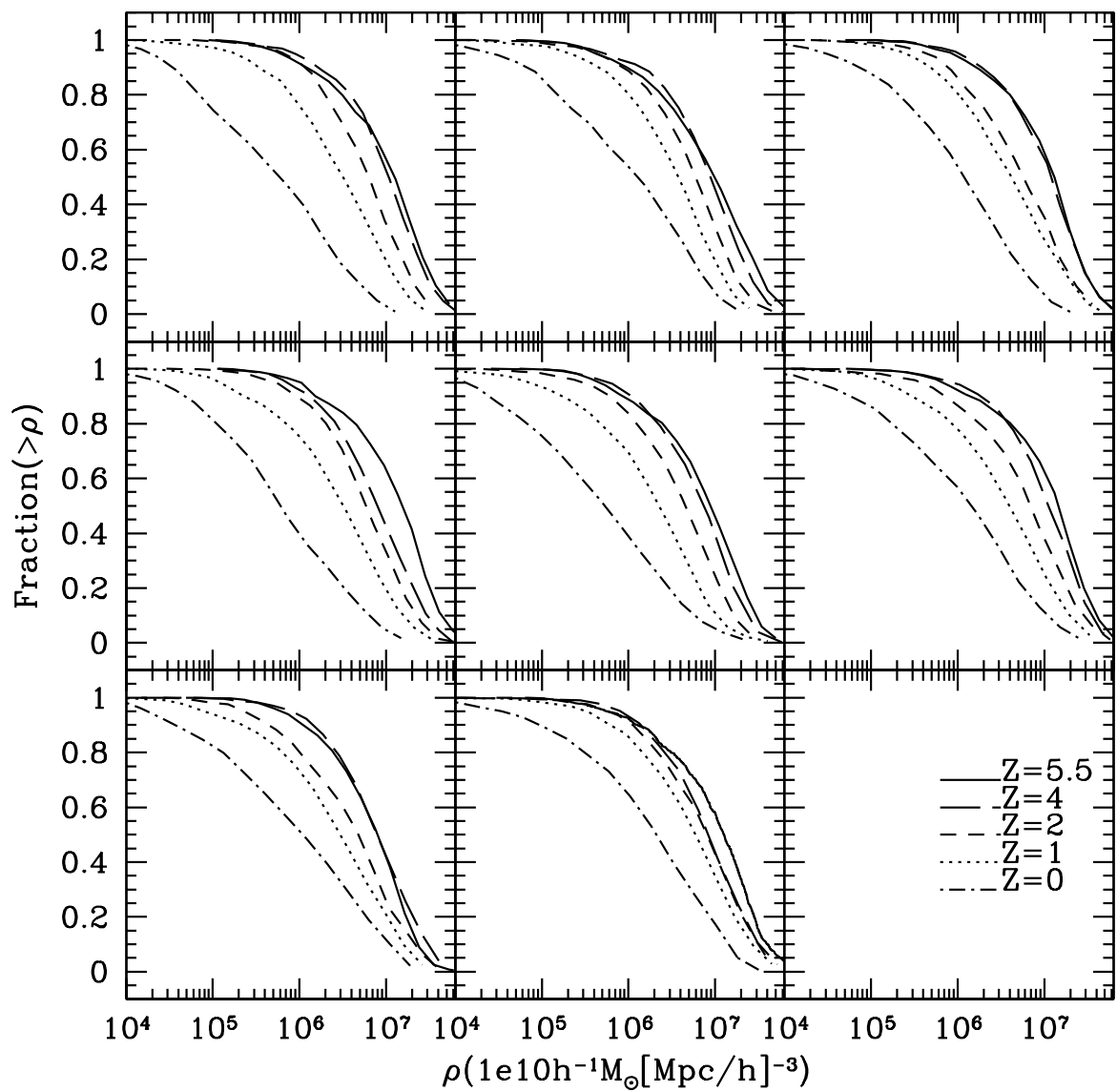} \caption{Evolution of the cumulative distribution
of ambient physical density for those 1000 particles within $r_{200}$ 
at $z=0$ which had the highest ambient densities at $z=6$.}
\label{fig-8}
\end{figure}

\begin{figure}[htbp]
\plotone{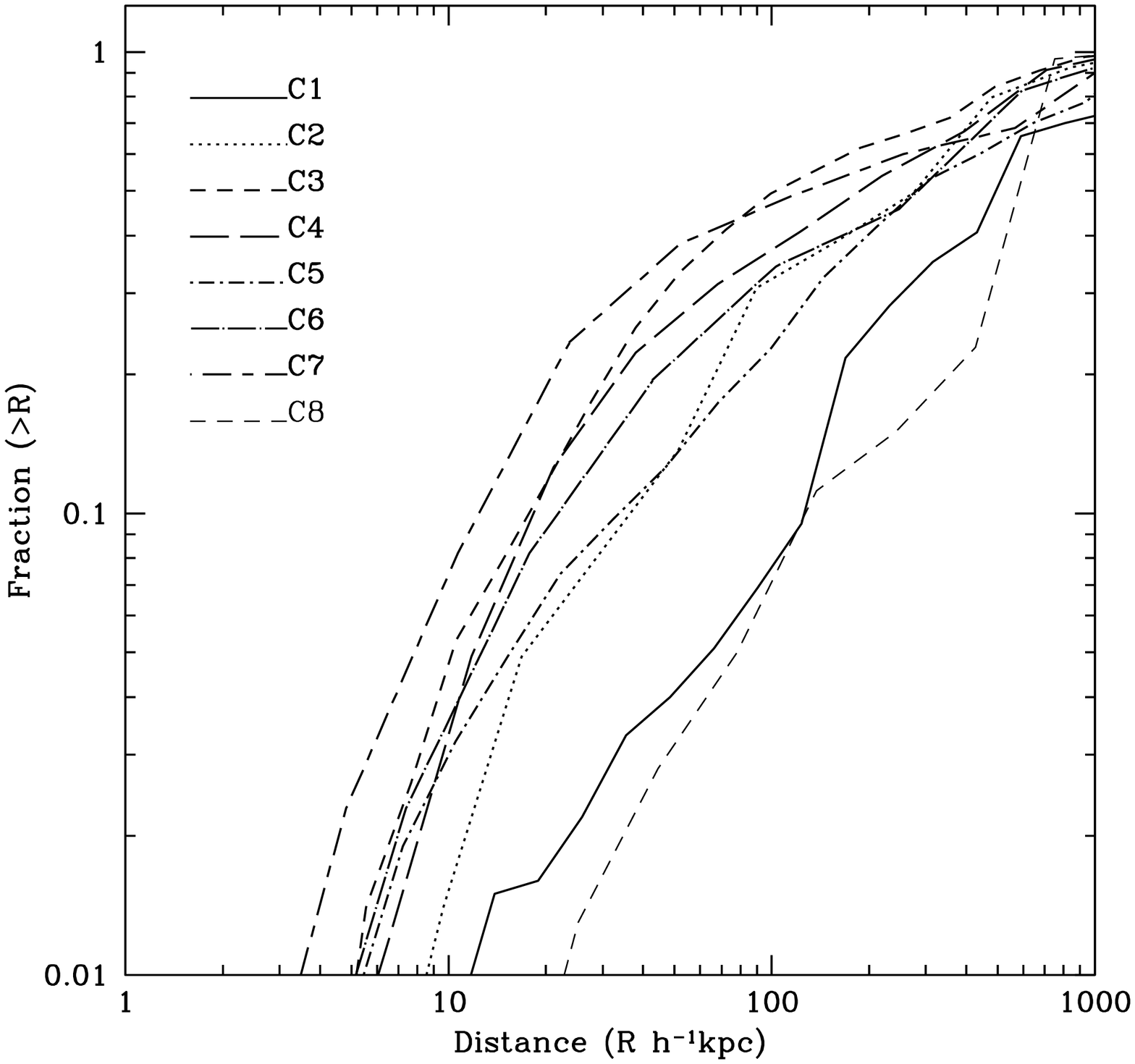} \caption{The cumulative radial distribution at $z=0$
of the particle sets tracked in Figure~8. The center used here is the
density center of the main halo. Note that in several cases a large
fraction of the particles are concentrated to one of the halo
substructures rather than to this main center.  } \label{fig-9}
\end{figure}

\subsection{The Attractor Hypothesis}

Our interpretation depends on the hypothesis of a dynamical attractor effect,
that {\it the inner cores of galaxies tend to approach, through multiple
mergers, a universal density profile for their collisionless mixture of stars
and dark matter}. The famous paper of Lynden-Bell (1967) introduced the idea
that violent relaxation of a collisionless gravitating system may drive it
towards a universal structure in the same way that a classical gas in a box
evolves from arbitrary initial conditions towards a spatially uniform
Maxwell-Boltzmann distribution; the coarse-grained structure of the system may
evolve towards an attractor. The NFW papers demonstrated behavior of this
kind, apparent evolution towards a universal density profile, in hierarchical
structure formation from cosmological initial conditions. Such universality
clearly requires the stability of the NFW mass distribution under violent
evolution through mergers, and some observational constraints suggest that
such an attractor is active in reality.

In the absence of the attractor effect the dissipative settling needed to
increase the baryon to dark matter ratio in the luminous parts of a galaxy
would tend to make the central mass density run steeper than the NFW/Moore
model, leading to two problems. First, it would seem to produce an
unacceptably steep central mass density run (Jesseit, Naab \&\ Burkert 2002;
Sand, Treu \&\ Ellis, 2002; Sand et al. 2003, and references therein). Second,
it would likely produce too many galaxies with large velocity dispersions. On
the other hand, if all the collisionless matter in a typical giant elliptical,
dark plus stellar, were driven by mergers into a good approximation to an NFW
profile, it would certainly help relieve the apparent challenge to the
$\Lambda$CDM model from the central mass density run in clusters. Also, it
would allow consistency between the comoving number density of massive halos
as a function of velocity dispersion at radius $r_e\sim 10h^{-1}$~kpc) and
SDSS counts of galaxies as a function of stellar velocity dispersion (as
illustrated in Figure~2 of Loeb \& Peebles( 2003) using the data from Sheth et
al. (2003)). The attractor hypothesis was invoked by Loeb \& Peebles (2003)
specifically to help resolve these two observational issues.

A similar argument is given in a recent paper by El-Zant et al (2004). These
authors performed simulations of idealised NFW clusters in which the galaxies
are represented by a population of ``massive clumps''. As the clumps spiral to
cluster center, their energy losses cause the central NFW cusp to flatten in
agreement with an analytic model proposed by El-Zant, Shlosman \& Hoffman
(2001). This effect also occurs in our own simulations as massive substructure
clumps merge into the central region, but in our case the clumps are not rigid
and are progressively disrupted as they move in. Stellar galaxies {\it can}
clearly be disrupted in like manner, but they are expected to be more
resistant to stripping than their dark matter halos as a result of the
dissipative concentration which produced them (White \& Rees 1978). It is thus
unclear whether the El-Zant et al. approximation of ``unstrippable'' galaxies
is more or less realistic than our own neglect of baryonic gravity. The
main point, perhaps, in the current context is that both approaches suggest
that the total mass distribution remains NFW-like in the inner regions with
the dark matter distribution expanding to compensate for the central
concentration of the stars.

In the attractor hypothesis the usual correction for compression by stellar
mass added through dissipative settling would apply only to stars formed out
of baryons added after the last major merger. Figure~3 indicates that the
central mass distributions in massive galaxies typically have been rearranged
by at least one major merger at $z<2$.  Our hypothesis then requires that most
of their stars formed earlier than that.

Some elliptical galaxies at $z<1$ do show evidence for recent star formation
(J{\o}rgensen 1999; Trager et al. 2000; Menanteau, Abraham, \& Ellis 2001;
Fukugita 2003), which might be the result of accretion or of recycling of
matter shed by stars within the galaxy. The amount of mass added or rearranged
by recent star formation is generally thought to be only $\sim 10$--$20\%$,
however, and so is not likely to greatly disturb the attractor solution.

\section{Open Issues}

The idea that some large ellipticals formed by merging of
late-type galaxies has been under discussion for many years
(e.g. Toomre \&\ Toomre 1972; Ostriker 1980; Negroponte \& White 1983;
Schweizer 2000). Under the attractor hypothesis, an elliptical that
formed by the merger of gas-rich galaxies with a subsequent starburst,
perhaps the typical path in galaxy groups, might be expected to show
significant baryon loading effects on its total density run; an
elliptical that formed by mergers of less gas-rich early-type
galaxies, perhaps the more common pattern for cluster ellipticals,
would show fewer baryon loading effects.  We are not aware of
observational tests of this possible systematic difference between
$\rho (r)$ in field and cluster galaxies, although possibly related
differences are seen between the central density runs of bright and faint
ellipticals (Faber et al 1997) and between the core colors of cluster and 
field ellipticals (Menanteau, Abraham \& Ellis 2001).

A related issue is the meaning of the strikingly small differences
between the spectra and mass-to-light ratios of cluster and group
ellipticals, as illustrated by Hogg et al. (2003) and van Dokkum \&\
Ellis (2003). A detailed analysis of this effect within the
$\Lambda$CDM model would be challenging, and certainly desirable. A
first analysis by Kauffmann \& Charlot (1998a) shows qualitative
agreement with the data but a quantitative difference between cluster
and field which may be larger than observed.

There has been a long debate over observational constraints
on the formation timescale for the mass concentrations
corresponding to the luminous parts of present-day giant galaxies
(e.g. Peebles 1989; White 1989; and references therein). Radio galaxy
surveys provide convincing evidence for the presence of old massive
galaxies at redshifts $1<z<3$ (Lilly \& Longair 1984; Nolan et
al. 2003; Willott et al. 2003).  Massive high redshift protogalaxies
are likely hosts for the $\sim 10^9M_\odot$ black holes that power the
SDSS quasar population at $z\sim 6$ (Fan et al. 2003; Wyithe \& Loeb
2003). On the other hand, a number of recent attempts to measure the
evolution of the mean stellar density contributed by massive galaxies
have concluded that only half the current stars are present at $z\sim
1$ to 1.5 and only a tenth at $z\sim 3$ to 4 (Drory et al. 2003; Bell
et al. 2003; Dickinson et al. 2003; Stanford et al. 2003).  Estimating
the number density and stellar mass of giant galaxies at high redshift
poses a severe observational challenge, however, and the current
situation is confused. Thus Bell et al. (2003) find that the stellar
mass in their red sequence of galaxies has increased by a factor of
two since redshift $z=1$, but Pozzetti et al. (2003) see no
significant evolution of the stellar mass function of massive galaxies
over the same redshift interval.

There also is continuing debate over the relation of the observations to 
current theoretical models. Kauffmann \& Charlot (1998b) find a considerable
difference between the redshift distributions predicted for K-selected samples
by the assumption of pure luminosity evolution out to high redshift and by a
semianalytic $\Lambda$CDM model for galaxy formation. They conclude that the
observations favor the latter. Somerville et al. (2003) find much smaller
differences between their own versions of these two models, the predicted
redshift distributions differing insignificantly at $z<1.4$. At higher
redshift their hierarchical model predicts fewer galaxies than their pure
luminosity evolution model, with the observations lying between the two. Our
own analysis indicates that in the $\Lambda$CDM model the stellar mass in a
giant galaxy at $z=3$ could be as much as half the present value. This is
considerably less rapid evolution than is claimed by many authors, but is
significantly later assembly than pure luminosity evolution assumes.  We
emphasise that the $\Lambda$CDM model does produce enough massive objects at
early times to account for the highest redshift galaxy clusters, massive
galaxies and luminous quasars (Efstathiou \& Rees 1992, Mo \& White 2002). The
debate is whether current treatments of star and black hole formation
adequately represent the predictions of the $\Lambda$CDM cosmology, and, of
course, whether these predictions are compatible with the observed numbers of
massive objects at high redshift.

The numerical simulations used in this paper suggest
the typical cD galaxy has suffered significant merging events at
redshifts less than unity.  The cluster Abell 2199 (Minkowski 1961)
has long been considered a likely example of galaxies observed in the
act of merging, and the cluster C0337 at $z=0.59$ may be another case
(Nipoti et al. 2003).  The number of candidate merging systems of this
type is not large, however. It would be of considerable interest to
use numerical simulations to develop diagnostics of the appearance of
recently merged, massive, early-type galaxies. These could then be
used to check the high merger rate of the $\Lambda$CDM cosmology.

Our discussion highlights two systematic effects of purely
gravitational halo formation. First, the form for the halo density run
behaves as a dynamical attractor (Navarro, Frenk \& White 1997; Huss, Jain
\&\ Steinmetz 1999).  Second, the mass within a fixed physical radius
around the most massive halos evolves little with time after reaching
a density contrast on the order of 100 (Fukushige \& Makino 2001; Loeb
\& Peebles 2003). Both effects are supported by numerical simulations,
but have not been fully checked in the specific context of baryon
settling. Existing simulations of mergers of spirals embedded within
NFW-like halos do produce remnants whose inner regions are closer to
NFW than those of their progenitors, despite remaining dominated by
stars (Barnes 1992; Dubinski 1998). Further simulations would be
helpful, however, to check our attractor hypothesis, in particular
whether a halo which is compressed relative to NFW by baryon loading
relaxes back to NFW after a few major mergers.

Finally, we note that since the physics of pure gravitating systems
is simple, even if their behavior is complex, there may be an analytic
explanation of the systematics of halo formation discussed in this
paper. Possible approaches are discussed by Syer \& White (1998) and
Dekel, Devor \& Hetzroni (2003) among others, but a convincing
explanation remains elusive.

\acknowledgements

We have benefitted from discussions with Masataka Fukugita
and Jeremy Heyl.  This work was supported in part by NASA grant
 NAG 5-13292, and by NSF grants AST-0071019, AST-0204514 (for A. L.).


\begin{thebibliography}{}

\bibitem[Barnes (1992)]{1992ApJ...393..484B}
Barnes, J.~E., 1992, \apj, 393, 484

\bibitem[Baugh, Cole, Frenk, \& Lacey (1998)]{1998ApJ...498..504B} Baugh,
C.~M., Cole, S., Frenk, C.~S., \& Lacey, C.~G.\ 1998, \apj, 498, 504

\bibitem[Bell et al. (2003)]{astro-ph/0303394}
Bell, E.~F., Wolf, C., Meisenheimer, K., Rix, H., Borch, A., Dye, S.,
Kleinheinrich, M., \&\  McIntosh, D.~H.\ 2003, \apj, submitted; astro-ph/0303394 

\bibitem[Bernardi et al. (2003)]{2003AJ....125.1882B}
Bernardi, M.,  et al.\ 2003, AJ, 125, 1866B

\bibitem{} Dekel, A., Devor, J. \& Hetzroni, G. 2003 \mnras, 341, 326.

\bibitem{} Conselice, C.~J., Bershady, M~.A., Dickinson,
M. \&Papovich, C., AJ, 2003, 126, 1183C

\bibitem[Dickinson et al.(2003)]{2003ApJ...587...25D} Dickinson, M.,
Papovich, C., Ferguson,  H.~C.,  \&\ Budavari, T., 2003, \apj, 587, 25

\bibitem[Drory et al.(2003)]{2003ApJ...595..698D} Drory, N, Bender,
R., Feulner, G., Hopp, U., Maraston, C., Snigula, J., \&\ Hill, G.~J.,
2003, \apj, 595, 698

\bibitem[Dubinski (1998)]{1998ApJ...502..141D}
Dubinski, J. 1998, \apj, 502, 141

\bibitem[El_Zant at al. (2001)]{}
El-Zant, A.~A., Shlosman, I. \& Hoffman, Y., 2001, \apj, 560, 636

\bibitem[El_Zant et al. (2004)]{astro-ph/0309412}
El-Zant, A.~A., Hoffman, Y., Primack, J., Combes, F., Shlosman, I., 
2004, \apj, 607, L75

\bibitem{} Efstathiou, G.~P. \& Rees, M.~J. 1988, \mnras, 230, 5p
 
\bibitem{} Faber, S.~M., Tremaine, S.~D., Ajhar, E.~A., Byun, Y.-I.,
Dressler, A., Gebhardt, K., Grillmair, C., Kormendy, J., Lauer, T.~R.
\& Richstone, D. 1997, \aj, 114, 177

\bibitem[Fan et al.(2003)]{2003AJ....125.1649F} Fan, X.~et al.\ 2003, \aj,
125, 1649

\bibitem[Frenk et al.(1985)]{1985Natur.317..595F} Frenk, C.~S., White,
S.~D.~M, Efstathiou, G.~P. , \&\ Davis, M, 1985, Nature, 317, 595

\bibitem[Fukugita]{} Fukugita, M. 2003, submitted to ApJ

\bibitem[Fukushige \& Makino(2001)]{2001ApJ...557..533F} Fukushige, T.,~\&
Makino, J.\ 2001, \apj, 557, 533

\bibitem[Ghigna et al.(2000)]{2000ApJ...544..616G} Ghigna, S.,
Moore, B., Governato, F., Lake, G., Quinn, T., \& Stadel, J.\
2000, \apj, 544, 616

\bibitem[Hamilton(1985)]{1985ApJ...297..371H} Hamilton, D.\ 1985, \apj,
297, 371

\bibitem[Hogg et al. (2003)]{Hogg} Hogg, D. et al. 2003, ApJL, submitted;
astro-ph/0307336.

\bibitem[Huss et al. (1999)]{1999ApJ...517...64H} Huss, A.~B., Jain, B., \&\ Steinmetz, M., 1999, \apj, 517, 64

\bibitem[Jenkins et al.(2001)]{2001MNRAS.321..372J} Jenkins, A., Frenk,
C.~S., White, S.~D.~M., Colberg, J.~M., Cole, S., Evrard, A.~E., Couchman,
H.~M.~P., \& Yoshida, N.\ 2001, \mnras, 321, 372

\bibitem[Jesseit, Naab, \& Burkert(2002)]{2002ApJ...571L..89J} Jesseit, R.,
Naab, T., \& Burkert, A.\ 2002, \apjl, 571, L89

\bibitem[J{\o}rgensen(1999)]{1999MNRAS.306..607J} J{\o}rgensen, I.\ 1999,
\mnras, 306, 607

\bibitem[Kauffmann et al. (1998a)]{1998MNRAS.294..705K}
Kauffman, G. \&\ Charlot, S. 1998a, \mnras, 294, 705 

\bibitem[Kauffmann et al. (1998b)]{1998MNRAS.297L..23K}
Kauffman, G. \&\ Charlot, S. 1998b, \mnras, 297, L23 

\bibitem[Klypin, A.]{Klypin}
Klypin, A., Kravtsov, A. V., Bullock, J. S., \& Primack, J. R. 2001,
ApJ, 554, 903

\bibitem[Lilly \& Longair(1984)]{1984MNRAS.211..833L} Lilly, S.~J.~\& 
Longair, M.~S.\ 1984, \mnras, 211, 833 

\bibitem[Loeb \& Peebles(2003)]{2003ApJ...589...29L} Loeb, A.~\& Peebles,
P.~J.~E.\ 2003, \apj, 589, 29

\bibitem[Lynden-Bell (1967)]{}
Lynden-Bell, D., 1967, \mnras, 136, 101

\bibitem[Menanteau, Abraham, \& Ellis(2001)]{2001MNRAS.322....1M}
Menanteau, F., Abraham, R.~G., \& Ellis, R.~S.\ 2001, \mnras, 322, 1

\bibitem[Minkowski(1961)]{1961AJ.....66..558M} Minkowski, R.\ 1961, \aj, 66, 558

\bibitem[Moore et al.(1999)]{1999ApJ...524L..19M} Moore, B.,
Ghigna, S., Governato, F., Lake, G., Quinn, T., Stadel, J., \&
Tozzi, P.\ 1999, \apjl, 524, L19

\bibitem{} Mo, H.~J. \& White, S.~D.~M. 2002, \mnras, 336, 112

\bibitem[Navarro, Frenk, \& White(1997)]{1997ApJ...490..493N} Navarro,
J.~F., Frenk, C.~S., \& White, S.~D.~M.\ 1997, \apj, 490, 493 (NFW)

\bibitem{}  Navarro, J.~F., Hayashi, E., Power, C., Jenkins, A.~R.,
Frenk, C.~S., White, S.~D.~M., Springel, V., Stadel, J. \& Quinn, 
T.~R.\ 2003, \mnras, submitted (astro-ph/0311231)
   
\bibitem[Negroponte \& white (1983)]{1983MNRAS....201..401}
Negroponte, J. \& White, S.~D.~M.\ 1983, \mnras, 201, 401

\bibitem[Nipoti et al.(2003)]{2003MNRAS.344..748N} Nipoti, C., Stiavelli,
M., Ciotti, L., Treu, T., \& Rosati, P.\ 2003, \mnras, 344, 748

\bibitem[Nolan, Dunlop, Jimenez, \& Heavens(2003)]{2003MNRAS.341..464N} Nolan,
L.~A., Dunlop, J.~S., Jimenez, R., \& Heavens, A.~F.\ 2003, \mnras,
341, 464

\bibitem[Oke(1971)]{1971ApJ...170..193O} Oke, J.~B.\ 1971, \apj, 170, 193

\bibitem[Oke(1984)]{1984cgg..conf...99O} Oke, J.~B.\ 1984, ASSL Vol.~111:
Clusters and Groups of Galaxies, 99

\bibitem[Ostriker(1980)]{1980ComAp...8..177O} Ostriker, J.~P.\ 1980,
Comments on Astrophysics, 8, 177

\bibitem[Ostriker \& Tremaine (1975]{OT} Ostriker, J. P. \& Tremaine,
S. D. 1975, \apj, 202, L113

\bibitem[Peebles(1989)]{P89} Peebles, P. J. E. 1989, in The Epoch of Galaxy
Formation, eds. C. S. Frenk et al. NATO ASI series, vol 264, p. 1

\bibitem[Peebles(2002)]{2002nec..conf..351P} Peebles, P.~J.~E.\ 2002, ASP 
Conf.~Ser.~283: A New Era in Cosmology, 351 

\bibitem[Pozzetti et al.(2003)]{2003A&A...402..837P} Pozzetti, L.~et al.\ 
2003, \aap, 402, 837 

\bibitem[Romanowsky, A. J. et al. 2003]{Romanowsky} Romanowsky, A.~J.,
Douglas, N.~G., Arnaboldi, M., Kuijken, K., Merrifield, M.~R., Napolitano,
N.~R., Capaccioli, M., \& Freeman, K.~C.\ 2003, Science, 301, 1696

\bibitem[Sand, Treu, \& Ellis(2002)]{2002ApJ...574L.129S} Sand, D.~J.,
Treu, T., \& Ellis, R.~S.\ 2002, \apjl, 574, L129

\bibitem[Sand, D. J.]{Sand} Sand, D. J., Treu, T., Smith, G., \& Ellis,
R. S. 2003, ApJ, submitted; astro-ph/0309465

\bibitem[ Schweizer, F. 2000]{ Schweizer} Schweizer, F. 2000, Phil Trans R
Soc London, A 358, 2063

\bibitem[Sheth, Mo, \& Tormen(2001)]{2001MNRAS.323....1S} Sheth, R.~K., Mo,
H.~J., \& Tormen, G.\ 2001, \mnras, 323, 1 , 119

\bibitem[Sheth, R. K. \& Tormen, G. 1999]{st}
Sheth, R. K. \& Tormen, G. 1999, MNRAS, 308

\bibitem[Sheth, R. K. et al. 2003]{Sheth} Sheth, R. K. et al. 2003, ApJ, in
press; astro-ph/0303092

\bibitem[Somerville et al. (2003)]{astro-ph/0309067}
Somerville, R.~S. et al., 2003, \apjl, in
press; astro-ph/0309067

\bibitem[Springel, Yoshida \&White]{Springel}
Springel, V., Yoshida, N., \& White S.~D.~M, 2001, New Ast. 6, 79

\bibitem[Stanford et al. (2003)]{astro-ph/0310231}
Stanford, S.~A., Dickinson, M., Postman, M., Ferguson, H., Lucas, R.,
Conselice, C., Budavari, T., \& Somverville R., 2003, \apj, in press;
astro-ph/0310231

\bibitem[Syer \& White(1998)]{1998MNRAS.293..337S} Syer, D.~\& White,
S.~D.~M.\ 1998, \mnras, 293, 337

\bibitem[Toomre \& Toomre(1972)]{1972ApJ...178..623T} Toomre, A.~\& Toomre,
J.\ 1972, \apj, 178, 623

\bibitem[Trager, Faber, Worthey, \& Gonz{\'
a}lez(2000)]{2000AJ....119.1645T} Trager, S.~C., Faber, S.~M., Worthey, G.,
\& Gonz{\' a}lez, J.~J.\ 2000, \aj, 119, 1645

\bibitem[van Dokkum \& Ellis(2003)]{2003ApJ...592L..53V} van Dokkum,
P.~G.~\& Ellis, R.~S.\ 2003, \apjl, 592, L53

\bibitem[White (1980)]{1980MNRAS.191P...1W}
White, S.~D.~M., 1980, \mnras, 191, 1

\bibitem[White (1989)]{W89} White, S. D. M. 1989, in The Epoch of Galaxy
Formation, eds. C. S. Frenk et al. NATO ASI series, vol 264, p. 15

\bibitem[White \& Rees (1978)]{}
White, S.~D.~M. \& Rees, M.~J., 1978, \mnras, 183, 341

\bibitem[Willott, Rawlings, Jarvis, \& Blundell(2003)]{2003MNRAS.339..173W}
Willott, C.~J., Rawlings, S., Jarvis, M.~J., \& Blundell, K.~M.\ 2003,
\mnras, 339, 173

\bibitem[Wyithe, J. S. B., \& Loeb, A. 2003]{Wyithe}
Wyithe, J.~S.~B.~\& Loeb, A.\ 2003, \apj, 595, 614

\bibitem[Yoshida et al. (2001)]{Yoshida} Yoshida N., Sheth R.~K. \&\ 
Diaferio A., 2001, \mnras, 328, 669

\end{thebibliography}
\end{document}